\documentstyle[preprint,prl,aps]{revtex}

\begin{document}

\draft

\title{Decoherence without dissipation}

\author{G. W. Ford}

\address{Department of Physics, University of Michigan, \\
Ann Arbor, MI 48108-1120}

\author{R. F. O'Connell}

\address{Department of Physics and Astronomy, Louisiana State University,\\
Baton rouge, LA 70803-4001}

\date{\today }

\maketitle

\begin{abstract}
The prototypical Schr\"{o}dinger cat state, i.e., an initial
state
corresponding to two widely separated Gaussian wave packets, is
considered.
The decoherence time is calculated solely within the framework
of elementary
quantum mechanics and equilibrium statistical mechanics. This
is at variance
with common lore that irreversible coupling to a dissipative
environment is
the mechanism of decoherence. Here, we show that, on the
contrary,
decoherence can in fact occur at high temperature even for
vanishingly small
dissipation.
\end{abstract}


\narrowtext

Quantum teleportation \cite{zeilinger}, quantum information and computation 
\cite{bennett,quantum}, entangled states \cite{haroche},
Schr\"{o}dinger
cats \cite{zeilingera}, and the classical-quantum interface
\cite{tegmark}:
topics at forefront of research embracing quantum physics,
information
science and telecommunications and all depending on an
understanding of
decoherence \cite{giulini}, i.e., how a quantum interference
pattern is
destroyed. In an introduction to the contents of a recent book
devoted
wholly to this subject, Joos surveys the current situation and,
in
discussing the mechanism of decoherence, states that ``...
irreversible
coupling to [a dissipative] environment seems to have become
widely accepted
...''. Here, while we agree that coupling to the environment is
necessary to
establish thermal equilibrium, we show that at high temperature decoherence
occurs even for vanishingly small dissipation. The situation is
like that
for an ideal gas: collisions are necessary to bring the gas to
equilibrium
but do not appear in the equation of state, nor in the velocity
distribution.

Much of the discussion of decoherence \cite{caldeira,walls,savage,zurek} has
been in terms of the simple problem of a particle moving in one
dimension
that is placed in an initial superposition state
(Schr\"{o}dinger ``cat''
state) corresponding to two widely separated wave packets. The
motivation
for this choice is that it can be applied, say, to describe the
interference
pattern arising in Young's two-slit experiment\cite{caldeira}
or that
arising from a quantum measurement involving a pair of
``Gaussian slits'' 
\cite{feynman,ford}. Of primary interest is the question of the
classical-quantum interface, i.e., how the interference pattern
is destroyed
with the evolution of a classical state corresponding to two
separately
propagating packets. Decoherence refers to this destruction of
the
interference pattern and key questions are what is the origin
of decoherence
and what is the time scale for less of coherence. The
maintenance of
coherence is an essential element in quantum teleportation,
etc. Thus, an
understanding of all physical phenomena which can cause
decoherence is
essential. Our purpose here is to give an elementary
calculation showing
that at high temperature ($kT\gg \hbar \gamma $, where $\gamma
$ is the
dissipative decay rate) decoherence occurs in a very short time
that is,
contrary to widely held belief, independent of the strength of
coupling to
the environment. Our starting point is the prototypical
Schr\"{o}dinger cat
state, i.e., an initial state corresponding to two separated
Gaussian wave
packets. The corresponding wave function has the form
\begin{equation}
\psi (x,0)=\frac{1}{[2(1+e^{-\frac{d^{2}}{8\sigma
^{2}}})]^{1/2}}\left( 
\frac{\exp \{-\frac{(x-\frac{d}{2})^{2}}{4\sigma
^{2}}+i\frac{mv}{\hbar }x\}%
}{(2\pi \sigma ^{2})^{1/4}}+\frac{\exp
\{-\frac{(x+\frac{d}{2})^{2}}{4\sigma
^{2}}+i\frac{mv}{\hbar }x\}}{(2\pi \sigma ^{2})^{1/4}}\right)
,  \label{1}
\end{equation}
where $\sigma $ is the width of each packet, $d$ is the
separation between
the centers of the two packets and $v$ is the particle
velocity. Next, we
solve the free particle Schr\"{o}dinger equation, 
\begin{equation}
i\hbar \frac{\partial \psi }{\partial t}=-\frac{\hbar
^{2}}{2m}\frac{%
\partial ^{2}\psi }{\partial x^{2}},  \label{2}
\end{equation}
with this initial state. The general solution is
\cite{merzbacher} 
\begin{equation}
\psi (x,t)=\sqrt{\frac{m}{2\pi i\hbar t}}\int_{-\infty
}^{\infty }dx^{\prime
}\exp \{-\frac{m(x-x^{\prime })^{2}}{2i\hbar t}\}\psi (x,0). 
\label{3}
\end{equation}
Hence, using (\ref{1}) we obtain 
\begin{eqnarray}
\psi (x,t) &=&\frac{e^{i\frac{mv}{\hbar
}x-i\frac{mv^{2}t}{2\hbar }}}{%
[2(1+e^{-\frac{d^{2}}{8\sigma ^{2}}})]^{1/2}}\left(
\frac{1}{[2\pi (\sigma +%
\frac{i\hbar t}{2m\sigma })^{2}]^{1/4}}\exp \{-\frac{(x-\frac{d}{2}-vt)^{2}}{%
4\sigma ^{2}(1+\frac{i\hbar t}{2m\sigma ^{2}})}\}\right.  
\nonumber \\
&&\left. +\frac{1}{[2\pi (\sigma +\frac{i\hbar t}{2m\sigma
})^{2}]^{1/4}}%
\exp \{-\frac{(x+\frac{d}{2}-vt)^{2}}{4\sigma
^{2}(1+\frac{i\hbar t}{%
2m\sigma ^{2}})}\}\right) .  \label{4}
\end{eqnarray}
Hence, the probability distribution, $P(x,t)=\left| \psi
(x,t)\right| ^{2}$
is 
\begin{eqnarray}
P(x;t) &=&\frac{1}{2(1+e^{-\frac{d^{2}}{8\sigma
^{2}}})\sqrt{2\pi (\sigma
^{2}+\frac{\hbar ^{2}t^{2}}{4m^{2}\sigma ^{2}})}}\left( \exp
\{-\frac{(x-%
\frac{d}{2}-vt)^{2}}{2(\sigma ^{2}+\frac{\hbar
^{2}t^{2}}{4m^{2}\sigma ^{2}})%
}\}+\exp \{-\frac{(x+\frac{d}{2}-vt)^{2}}{2(\sigma
^{2}+\frac{\hbar ^{2}t^{2}%
}{4m^{2}\sigma ^{2}})}\}\right.   \nonumber \\
&&\left. +2\exp \{-\frac{(x-vt)^{2}+\frac{d^{2}}{4}}{2(\sigma
^{2}+\frac{%
\hbar ^{2}t^{2}}{4m^{2}\sigma ^{2}})}\}\cos \frac{\hbar
td(x-vt)}{4m\sigma
^{2}(\sigma ^{2}+\frac{\hbar ^{2}t^{2}}{4m^{2}\sigma
^{2}})}\right) .
\label{5}
\end{eqnarray}
This is all within the realm of conventional quantum mechanics
\cite{merzbacher}.

Next we consider the case of a particle in thermal equilibrium, but so
weakly coupled to the environment that we can neglect
dissipation. The
principles of statistical mechanics then tell us that we obtain
the
corresponding probability distribution by averaging the
distribution (\ref{5}%
) over a thermal distribution of velocities. The result is 
\begin{eqnarray}
P_{{\rm T}}(x;t) &\equiv &\sqrt{\frac{m}{2\pi kT}}\int_{-\infty
}^{\infty
}dv\exp \{-\frac{mv^{2}}{2kT}\}P(x;t)  \nonumber \\
&=&\frac{1}{2(1+e^{-\frac{d^{2}}{8\sigma ^{2}}})\sqrt{2\pi
w^{2}}}\left(
\exp \{-\frac{(x-\frac{d}{2})^{2}}{2w^{2}}\}+\exp
\{-\frac{(x+\frac{d}{2}%
)^{2}}{2w^{2}}\}\right.   \nonumber \\
&&\left. +2\exp
\{-\frac{x^{2}}{2w^{2}}-\frac{w^{2}+\frac{kT}{m}t^{2}(\frac{%
\hbar t}{2m\sigma ^{2}})^{2}}{(\sigma ^{2}+\frac{\hbar
^{2}t^{2}}{%
4m^{2}\sigma ^{2}})w^{2}}\frac{d^{2}}{8}\}\cos \frac{\hbar
tdx}{4m\sigma
^{2}w^{2}}\right) ,  \label{6}
\end{eqnarray}
where we have used a subscript ${\rm T}$ to emphasize that this
is the
probability distribution at finite temperature and where we
have introduced 
\begin{equation}
w^{2}(t)=\sigma ^{2}+\frac{kT}{m}t^{2}+\frac{\hbar
^{2}}{4m^{2}\sigma ^{2}}%
t^{2}.  \label{7}
\end{equation}
This probability distribution is the sum of three contributions,
corresponding to the three terms within the parentheses. The
first two
clearly correspond to a pair of separately expanding wave
packet, with $%
w^{2}(t)$ the width of each, while the third term, the one
involving the
cosine, is an interference term. The attenuation coefficient
$a(t)$ is the
ratio of the factor multiplying the cosine to twice the
geometric mean of
the first two terms. Thus 
\begin{eqnarray}
a(t) &=&\exp \{-\frac{kTd^{2}}{8m\sigma ^{2}w^{2}}t^{2}\} 
\nonumber \\
&=&\exp \{-\frac{\frac{kT}{m}t^{2}d^{2}}{8\sigma ^{4}+8\sigma
^{2}\frac{kT}{m%
}t^{2}+\frac{2\hbar ^{2}t^{2}}{m^{2}}}\}.  \label{8}
\end{eqnarray}

At $t=0$, we see that $a(0)=1$ corresponding to maximum coherence and, as
mentioned above, the goal of experimentalists is to maintain
this coherence.
However, for very short times, we see that $a(t)\cong \exp
\{-t^{2}/\tau
_{d}^{2}\}$, where the decoherence time is 
\begin{equation}
\tau _{d}=\frac{\sqrt{8}\sigma ^{2}}{\bar{v}d},  \label{9}
\end{equation}
in which $\bar{v}=\sqrt{kT/m}$ is the mean thermal velocity.
This
decoherence time is much different from that quoted extensively
in the
literature\cite{giulini,zurek}, namely $\gamma ^{-1}\hbar
^{2}/mkTd^{2}$,
which is inversely proportional to $\gamma $ (the dissipative
decay rate).
By contrast, $\tau _{d}$ given by (\ref{9}) is independent of
$\gamma $. The
reason why existing calculations fail to obtain the form
(\ref{8}) for the
attenuation coefficient at short times is that they are based
on the
assumption that the initial state of the particle is a pure
state (of the
form (\ref{1}) with $v=0$) and use a master equation to
describe the time
development \cite{giulini,caldeira,walls,savage,zurek}. Such a
pure state is
effectively at zero temperature and when the particle is
suddenly coupled to
a bath at temperature $T$, as described by the master equation,
it takes a
time of order $\gamma ^{-1}$ for the particle to warm up and
acquire a
thermal distribution of velocities. Such an approach therefore
misses the
initial thermal distribution of velocities responsible for the
rapid loss of
coherence we have obtained. The result we have obtained does
follow from a
new approach \cite{ford}, which is not based on a master
equation and which
incorporates both arbitrary temperature and arbitrary
dissipation. In fact,
the exact general formula for the attenuation coefficient,
expressed in
terms of the mean square displacement and the nonequal-time
commutator, is
given by \cite{ford} 
\begin{equation}
a_{{\rm exact}}(t)=\exp \{-{\frac{s(t)d^{2}}{8\sigma
^{2}w_{{\rm exact}%
}^{2}(t)}}\},  \label{10}
\end{equation}
where now the width of a single wave packet is given by 
\begin{equation}
w_{{\rm exact}}^{2}(t)=\sigma
^{2}-{\frac{[x(t_{1}),x(t_{1}+t)]^{2}}{4\sigma
^{2}}}+s(t),  \label{11}
\end{equation}
in which 
\begin{equation}
s(t)=\langle \{x(t_{1})-x(t_{1}+t)\}^{2}\rangle ,  \label{12}
\end{equation}
is the mean square displacement. For the special case of a free
particle
without dissipation, where $s(t)=(kT/m)t^{2}$ and
$[x(t_{1}),~x(t_{1}+t)]=i{%
\hbar }t/m$, this reduces to (\ref{8}) above. However, even in
the presence
of dissipation, for times short compared with $\gamma ^{-1}$,
where $\gamma $
is a typical dissipative decay rate, the motion is again that
of a free
particle. For such free motion, there is a rapid decay of
coherence with
characteristic time $\tau _{d}$ given by (\ref{9}). It should
be stressed
that since decoherence decay times are always much smaller than
dissipative
decay times, we always have $\gamma \tau _{d}<<1$. For example,
if we
consider an electron at room temperature (300K), then
$\bar{v}=6.{8}\times
10^{6}cm/s$ so that if we take $d=1cm$ and $\sigma =0.{4}{\AA
}$, then using
(\ref{9}) we obtain $\tau _{d}=6.{9}\times 10^{-24}s$, which is
orders of
magnitude smaller than typical $\gamma ^{-1}$ values. Even for
$T=1K$ (which
fulfils our definition of high temperature, i.e., $kT\gg \hbar
\gamma $, for 
$\gamma \ll 10^{11}s^{-1}$) we obtain $\tau _{d}=1.{2}\times
{10}^{-22}s$.

In summary, we have presented a simple derivation of the result for
decoherence without dissipation, working solely within the
framework of
elementary quantum mechanics and equilibrium statistical
mechanics.

\end{document}